# On the use of STM superconducting tips at very low temperatures

J.G. Rodrigo[1], H. Suderow and S. Vieira

*Laboratorio de Bajas Temperaturas,*
*Departamento de Física de la Materia Condensada,*
*Instituto de Ciencia de Materiales Nicolás Cabrera,*
*Facultad de Ciencias*
*Universidad Autónoma de Madrid, 28049 Madrid, Spain*

*May 12, 2004*

## Abstract

We report on high quality local tunnel spectroscopy measurements in superconductors using in-situ fabricated superconducting tips as counterelectrode. The experiments were made at very low temperatures using a dilution refrigerator and a $^3$He cryostat. Spectra obtained with superconducting tip and sample of Al show that the spectroscopic resolution of our set-up is of 15µeV . Following the observation of Josephson current in tunneling regime (with tips of Pb and of Al), we discuss the feasibility of Scanning Josephson Spectroscopy with atomic size resolution. Experiments showing new applications of these superconducting tips under applied external magnetic fields are also reported.

## 1 Introduction

As it is well known, scanning tunnelling microscopy (STM) has grown to be, in the very short time of twenty years, one the most important tools towards our knowledge and control of the nanoworld. A pointed electrode, the tip, scans conducting surfaces closely, so that tunneling currents in the range of the nanoamperes can flow between tip and sample. The relevance of a good control over the tip characteristics was soon realized. Sharpness and cleanness are of paramount importance to guarantee spatial resolution and useful spectroscopic information. The combination of STM with ultrahigh vacuum techniques, UHV, has made possible to achieve these goals. But the choice of materials is limited by many special requirements. The wide spread of the STM between experimentalists of different fields has produced a big amount of results, many of them having as a goal to get information about the gap that opens at the Femi level in the density of states of superconductors. As it was shown by Giaever[1], forty years ago, tunnel I-V characteristics of a normal-insulatorsuperconductor, NIS, junction, contain this information. Most of these studies require working at liquid helium temperatures. As discussed by us in previous works[2-4], in addition to a well characterized tip, other experimental aspects, mainly regarding electromagnetic filtering, have to be carefully considered to get precise information about the electronic density of states. Unfortunately the literature is full of curves reported as superconducting gaps, whose aspect is far from the curves normally found with planar tunnel junctions. Rounded or missing coherence peaks, non-thermal finite conductance inside the gap region, and other non-intrinsic features, have been often published. Obviously the shape of the I-V curves, if properly measured, is of extreme importance in order to do a meaningful comparison with the current theoretical model predictions about the superconducting order parameter. In the following we are going to address our point of view to different aspects related to tunnelling measurements made at low temperatures in superconductors with the STM. We focus on novel improvements based on tips whose superconducting properties can be easily tuned at will, and we present the interesting advances which can be made in the characterization of superconducting materials by using them. Our approach relies on the unprecedented control on the minimal distance between macroscopic size objects, and on the fabrication of atomic size metallic contacts, that the STM has introduced. In the last ten years several groups have made possible, taking advantage of this capability, experiments going from the jump to the one atom contact, towards the evolution of connective necks of very small sizes [5–8]. The observed conductance steps and its relationship to the quantization of the conductance have triggered a considerable number of experimental and theoretical works. In some superconducting elements, as Pb, Nb or Al, measured at liquid helium temperatures, it has been possible, using the highly non linear form of the subgap conductance due to Andreev reflection processes, to relate the electronic conduction through a one atom contact to the number of transmitting channels[9]. This number turns out to be characteristic of the electronic structure of the connecting atom, in agreement with previous experiments and calculations for several normal metals [10]. A breakthrough in the world of nano-engineering has been the construction of chains of atoms of gold and the characterization of its mechanical and electrical properties [11,12]. The forces involved in the process of formation and the dissipation mechanisms have been investigated in detail in this unique one-dimensional system. Recently, atomic chains of gold atoms connecting superconducting electrodes were also created and studied [13]. In this case the electrodes were superconducting lead, on top of which a thin film (of thickness 20nm) of gold was evaporated, where superconductivity was induced by the proximity effect.

---

[1] E-mail: jose.rodrigo@uam.es

Larger nanostructures can also be created using the STM[14–16]. When the tip is deeply indented into the sample, an intimate contact between both electrodes is established. The region of the contact, called connective neck, can be elongated at will by means of repeated back and forth movements of the driving piezo while the tip is receded from the sample. This process results in the creation of a nanobridge between the electrodes, which can be broken apart leading to two nanostructures, the nanotips. The materials that we use to produce this structures, lead and aluminum, are very ductile, which permits to obtain easily nanobridges as long as several hundreds of nanometers. Both elements transit to a superconducting state at low temperatures, and due to the reduced atomic mobility at very low temperatures the nanostructures, including the apex which can be of atomic size, do not change its shape with time. This allows the use of these tips as atomic size superconducting probes to study a variety of complex systems. They are cone-like shaped, with lengths between 10 and 200 nm and the opening angles range from 60 to 10 degrees. It has been shown [17–20] that, under magnetic fields higher that the bulk critical one, the condensate becomes confined in the tip apical region.

We report in this article on the fabrication and characterization at low temperatures of STM tips of lead and aluminum. In the superconducting state both elements represent two different cases, due to different electron-phonon coupling intensity, well understood in the framework of BCS theory and its extension due to Eliashberg. We also discuss the application of these tips to investigate several superconducting properties, at very low temperatures and under applied magnetic fields. We present first the experimental setup, paying special attention to the requirements of resolution and stability needed to perform the different studies, and discuss subsequently the characteristics of the superconducting density of states at the nanostructure. Several examples of the applications of the superconducting nanotip to the study of different transport phenomena and different materials are finally shown.

**2 Spectroscopic resolution**

**2.1 S-S and S-N tunnel junctions at very low temperatures.**

The experiments on superconducting materials and nanostructures are currently performed in our group using several similar home built low temperature STM units. One of these can be cooled down to 70 mK, our lower limit in temperature, in a dilution refrigerator which was partially home built. Actually, the mixing chamber goes down to 25 mK, but there is a non-negligible difference with the STM temperature due to a necessarily long thermal path to the sample and tip holders. We have tested in this system the spectroscopic resolution of our setup (electronics and filtering arrangements). For this purpose we have measured the I-V tunnel characteristics of aluminum in the superconducting state. Aluminum is possibly one the best known superconducting materials, with a critical temperature of 1.2K and weak electron-phonon coupling. We fabricate the superconductor-superconductor, S-S, tunnelling junction from the controlled rupture of nanobridge, as previously described. Ideal I-V curves in the S-S tunnelling regime at very low temperature present the well known features of zero current up to the gap edge at $2\Delta$, where there is a jump to non-zero current[21,22]. This shows in the conductance curves (*dI/dV* vs *V*) as a divergence at energy $2\Delta$, (present at all temperatures) which is the sharpest feature that can be observed in tunneling spectroscopy measurements in superconductors.

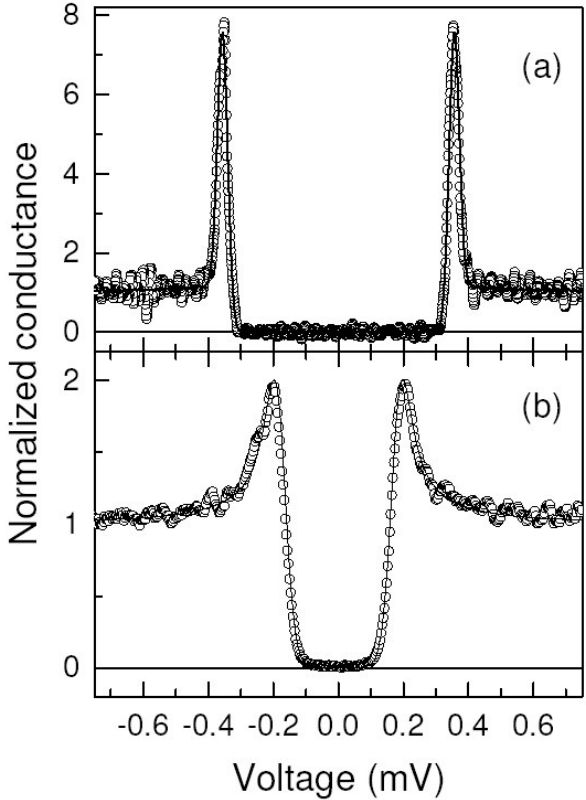

**Fig. 1**.
Normalized conductance curves obtained in tunneling regime when both tip and sample are made of Al (a), and when the Al tip is located over a gold sample (b). The experimental curves (circles) were obtained at the lowest temperature of the system (70 mK, see text) and for $R_N$ = 1$M\Omega$. Theoretical curves (lines) were calculated as described in the text, using the values T=0.150K and $\Delta$=175$\mu$eV.

Neglecting gap anisotropy or strong coupling effects (which is a very good approximation in Al), the only limit is set by the actual resolution in voltage (or energy) of the spectrometer. Therefore, this measurement is a direct test of the experimental set-up. The observed width of these features will be taken as a measure of the resolution in energy of our spectrometer. This energy resolution can be introduced in the calculus of the curves as a narrow gaussian distribution, which simulates the noise in the voltage source, and has a halfwidth in energy of $\sigma$. Note that this procedure is radically different from the lifetime smearing effect introduced in [23] and used in many spectroscopic experiments.

In Fig.1a we present an experimental S-S conductance curve and the corresponding fitting. The calculated curve was obtained with the parameters $\Delta = 175\mu eV$, $T = 70mK$ (base temperature of the system) and energy halfwidth $\sigma = 15\mu eV$. Actually, at non-zero temperature the current within the gap expected for these junctions is not zero due to the thermal broadening of the Fermi edge. However, at low temperatures, this current disappears exponentially and it is hardly detectable. Therefore, within our experimental resolution in current (1pA), we obtain the same curve up to 250mK. Once the resolution in energy known, it is possible to obtain the actual temperature of the tunnel junction, through the analysis of N-S curves. In this case there is no discontinuity in the current at the gap energy in the I-V curves, which show neatly the rounding effect of temperature. In fig.1b we present a N-S experimental conductance curve obtained after displacing the same Al tip to a sample of gold. This curve is fitted using conventional isotropic BCS s-wave theory and the energy distribution previously obtained for the S-S case, giving $\Delta = 175\mu eV$ and $T = 150mK$, which is the actual temperature achieved in the junction region. We want to stress that this precise determination of the spectroscopic resolution of the system is possible only through the S-S curves, and that the determination of the experimental temperature of the tunnel junction should be done afterwards, from the analysis of the N-S curves.

This process is possible thanks to the method of preparation of the superconducting tips and the displacement capabilities of our system. Indeed, an important aspect of our STM setup is the high capability of controlled displacement of the sample holder in the *xy* plane. As reported elsewhere[24], this feature permits the tip fabrication and ulterior use without any change in temperature or other ambiance conditions. These processes can be repeated at will if there is any suspicion on the tip sharpness or cleanness. As a matter of fact, we routinely test that the work function is of several eV before doing the spectra. We stress that our resolution is similar to the one achieved in many planar junction experiments[21], with the important differences that in STM measurements the tunnelling current is local and usually 4 to 6 orders of magnitude smaller, and that electrons tunnel through the vacuum, without any additional insulating barrier. The obtention of the above mentioned spectroscopic capabilities does not impair the imaging performances of the system. As an example of the scanning capabilities, and of the atomic sharpness of the superconducting nanotips created following our method, we present in the inset of fig.2 an image with atomic resolution of an area of $NbSe_2$ scanned with a Pb tip at 0.3K.

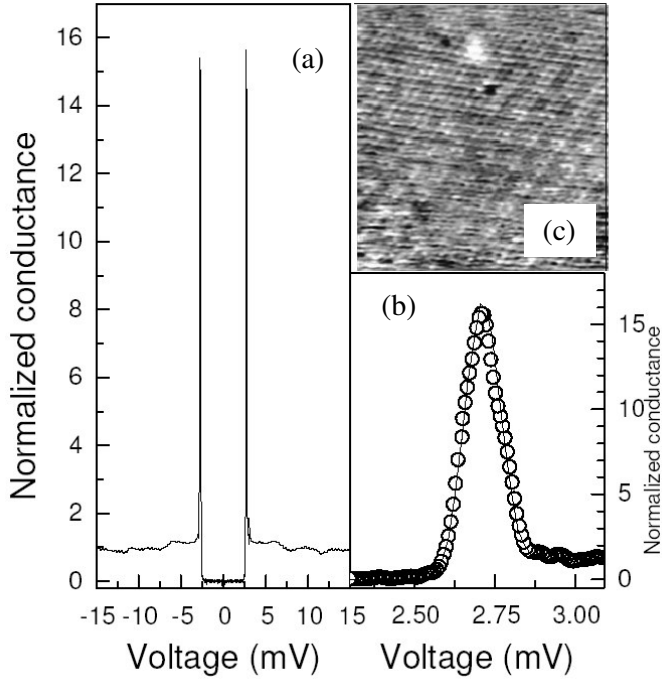

**Fig. 2.**
(a) Tunneling conductance curve obtained in a $^3$He cryostat at 0.3K with tip and sample of Pb ($R_N$ = 1MΩ). The gap edge is zoomed in (b). Theoretical conductance (line) has been calculated with the parameters described in the text, in order to reproduce the experimental curve (circles). The image (c) shows an area of $NbSe_2$ scanned with atomic resolution using a Pb tip at 0.3K.

**2.2 Determination of the gap distribution in lead, by S-S spectroscopy.**

As it is mentioned in the introduction, one of the most important aspects to be taken into account in tunnelling spectroscopy experiments concerns the quality of the spectra. This subject involves a series of important requirements in the experimental setup, stated above, and an exact knowledge of the electronic properties of the superconducting nanotip, which eventually will be used to investigate different phenomena and materials. The method of creation of the nanotip that we have developed, in situ at low temperatures, ensures that the tip apex will be clean, free from oxides or any type of external contamination. This fact leads to the observation of spectra like the one obtained for aluminum (Fig.1), in which a perfect agreement with the BCS theory is found. This is a very important result, showing that a nanoscopic superconducting tip, at zero magnetic field, presents the density of states predicted by theory for bulk superconductors. A strong coupling superconductor (Pb) can be used also to create a nanotip. The spectrum obtained in this case, with our STM mounted in a $^3$He cryostat, shows the expected signature of the main phonon modes, which has been followed as a function of the magnetic field in ref.[25], as well a quite broadened gap edge (fig. 2). This broadening is due the anisotropy of gap values in lead, which was already discussed in early tunnel junction experiments[26–29]. The width of the superconducting gap in lead (i.e., the distribution of gap values) measured in our experiment be obtained after taking into account the determination the energy resolution stated above. The temperature 0.35K) and energy resolution ($\sigma = 20\mu eV$) are fixed, we leave the superconducting gap $\Delta$ and the halfwidth of the distribution of values of the superconducting gap $\delta$ as free parameters, obtaining $\Delta = 1.35 meV$ and $\delta = 25\mu eV$.

We remark that only the precise determination of our limit resolution, which follows from our method of in situ fabrication of superconducting tips, permits to extract relevant information, as the one about the gap distribution lead, from local tunneling experiments. Recently there have been several reports on new superconducting materials which indicate that a single gap in the Fermi surface not the more frequent case [2,30–33]. Multiband superconductivity, and gap anisotropy, seem to be more ubiquitous than previously thought. We claim that this enhances importance of precise local tunnel measurements to shed light on the opened problems. The study of the order parameter close to $T_c$ in order determine the presence of asymmetric or multiband superconductivity in materials like NbSe$_2$ has been possible by the use of a superconducting Pb tip [32], due to the enhancement of the gap related features resulting from convolution of two superconducting DOS (tip and sample). This enhancement also improves the study of materials with very low $T_c$ (and gap values)[34].

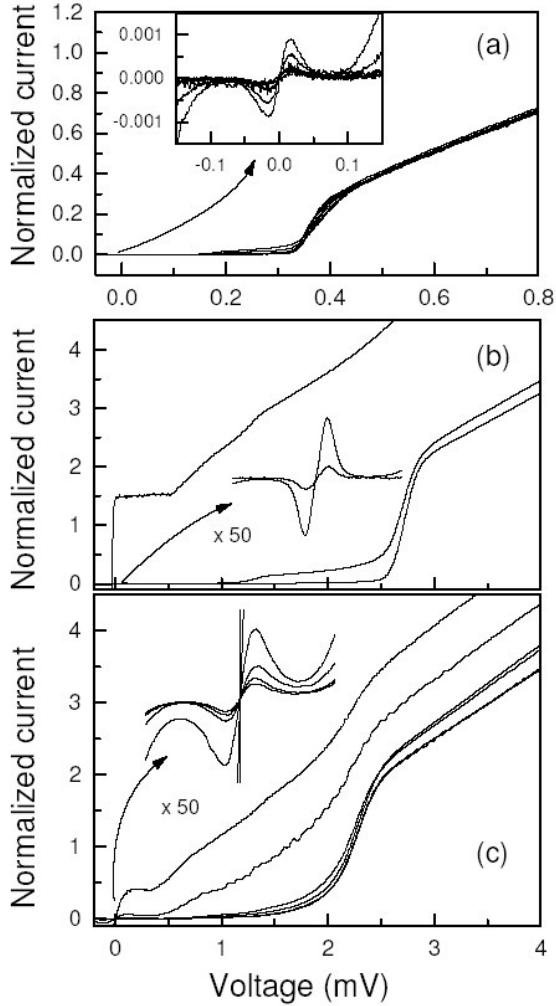

**Fig. 3**.
Normalized I-V curves ($IR_N$ vs $V$) corresponding to three different juctions:
(a) Al-Al, at $T = 150mK$, and $R_N$ from 3 to 0.16 MΩ;
(b) Pb-Pb, at $T = 0.35K$, and $R_N$ = 100, 20 and 0.2 kΩ;
(c) Pb-NbSe$_2$, at $T = 0.35K$, and $R_N$ from 100 kΩ down to 100Ω.
The curves with resistances below 1kΩ, taken at 0.3K, for which $k_BT$ is clearly smaller than $E_J$, are not in the dissipative regime.

### 2.3 The feasibility of Scanning Josephson spectroscopy

The precise knowledge that we achieve about the electronic density of states, DOS, of the tip allows a straightforward application to the study of different phenomena. One of these applications is Local Josephson Tunnelling Spectroscopy with atomic spatial resolution. As noted in refs.[35–39], the measurement of the Josephson current of Cooper pairs in atomic size and high resistance vacuum junctions, as the one that we deal with the STM experiments, is a true challenge.

Josephson binding energy, which defines the energy scale for the phase coupling Josephson junctions, is given by $E_J = \Delta R_Q /2R_N$, where is the superconducting gap, $R_N$ the normal state resistance of the junction and $R_Q = h/4e = 6.45k\Omega$. The thermal energy, $k_BT$ is higher that $E_J$ for normal resistances in the MΩ range, and not very low temperatures. For Pb junctions, both energies are similar at 50 mK, normal resistance of 1 MΩ. As discussed in refs.[36,37], for thermal energies bigger than, but comparable to Josephson binding energy, pair tunnelling would be observed, but the pair current will be dissipative, i.e. with the voltage drop proportional to the rate of thermally induced phase slips across the junction. By reducing distance between tip and sample, it is possible to cover a wide range of resistance and temperatures[36,37], which can permit to get information on the different Josephson regimes by changing in a controlled way the ratio between thermal and Josephson binding energies. In fig.3 we show the I-V curves, close to zero bias, of three different juctions Al-Al, Pb-Pb and Pb-NbSe$_2$, obtained in dissipative regime conditions. Also we show for Pb-Pb junctions I-V curves taken at 0.3K, with resistances below 1kΩ, where $k_BT$ is clearly smaller than $E_J$. A detailed analysis of these experiments will be reported elsewhere.

## 3 On the behavior of superconducting tips under magnetic fields

As has been show in previous reports[17,25,18–20,40,41], following to the application of an external magnetic field, higher than the critical one, the superconducting condensate becomes confined in a region around the tip apex. Taking into account the conelike shape of the tip, its length and opening angle determines, using single models, its superconducting properties. Transport measurements in tunnel and atomic size contact regimes fit very well the theoretical predictions. Due to the method used to create the tip, it is possible to build it up searching for predetermined properties. It is easy to understand that small opening angles and lengths, favor stronger response against magnetic fields. We have discussed this point in a previous article, where we studied the behavior of a tip of lead, with a low temperature critical field 40 times higher than the bulk critical one[20]. This kind of superconducting nanostructure is unique in several aspects. The first is that its connection to the normal region is perfect, being possible to modify the phase boundary characteristics, and the superconducting properties for a given nanostructure, by changing magnetic field and temperature. The second is that the information about its properties, comes from the tunnelling density of states, i.e. an equilibrium property. The third is that, by driving the tip into the contact regime, and inducing small changes of the minimal cross-section, it is possible to pass very large currents (up to mA) through the nanosized superconducting region. A non-equilibrium transport regime is attained, in which phase slip phenomena and large contact resistances are observed [40]. Many possible experiments could be thought in which the discussed nanostructure became useful.

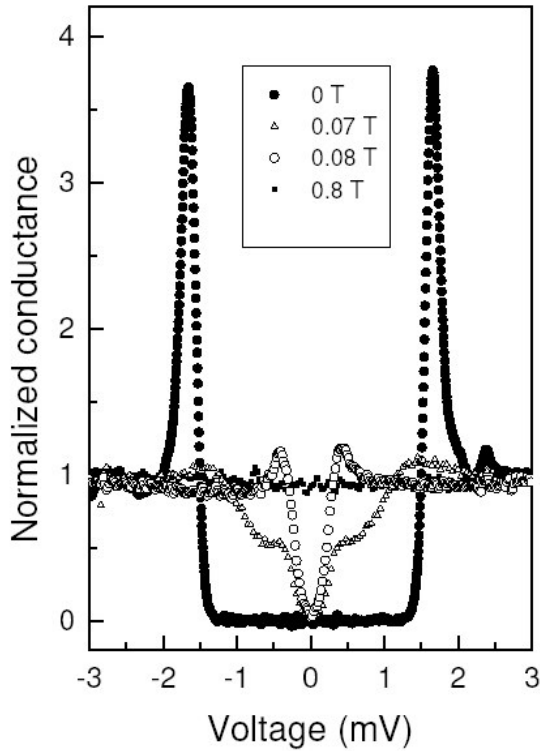

**Fig. 4**.
Normalized tunnelling spectra obtained using a superconducting Pb tip on a (NbSe$_2$)(LaSe)$_{1.14}$ sample (from [34]). At zero field (filled circle), the gap edge appears at $\Delta_{tip} + \Delta_{sample}$, characteristic of S-S' junctions. When the applied magnetic field is just above the bulk critical field of Pb (0.07$T$ at 0.35$K$), superconductivity in the tip becomes gapless, and a gap structure due to the sample appears at zero bias (triangles). At higher field superconductivity in the nanotip disappears revealing clearly the density of states of the sample (open circles), whose critical field is much larger (filled squares).

A simple one, done in the tunneling regime where tip and sample are separated through the vacuum barrier is shown in fig.4. A previously characterized tip of lead is moved with the xy table to probe another superconductor (NbSe$_2$)(LaSe)$_{1.14}$, whose tunnelling spectroscopy was never measured before[34]. By increasing the applied magnetic field, $H$, it is possible to test the density of states of the sample with three "different" tips: fully superconducting for fields below the bulk critical field of the tip $H_{C,bt}$, superconducting only in the apex region when $H_{C,bt} < H < H_{C,nt}$ ($H_{C,nt}$ is the critical field of the nanotip), and normal for $H > H_{C,nt}$. In this experiment, at 0.35K, the different critical fields are $H_{C,bt}$ =

0.07$T$, $H_{C,nt}$ = 0.08$T$, and $H_{C,sample}$ = 0.8$T$. Due to the very different values of the upper critical field, we can readily assume that the sample density of states does not show any significant changes during the whole procedure. In the lowest field curves, features characteristic of S-S' junctions show up and permit to obtain a precise value for the (a priori unknown) superconducting gap of the sample. This value is used to do a fit to the BCS model to the curves corresponding to higher fields. Deviation from this model would be apparent if we would deal with non-conventional superconductors. Above the bulk critical field of lead, the tip apex becomes gapless. Note that, in this situation, the density of states of the tip will be sensitive to stray fields coming from the sample, as the ones caused by vortices. Currently, we are further developing the use of such tips, as other kinds of local magnetic inhomogeneities in other samples may be easily imaged with them.

**4 Summary**

We have reported in this article on the fabrication and use in situ, for several purposes, of well characterized atomic size superconducting tips of lead and aluminum. The experiments were performed at very low temperatures, and special cautions were taken to improve the electromagnetic ambiance in the measurement region. We have determined the experimental resolution of our spectrometer (15$\mu eV$), and given neat evidences about the interesting possibilities of our technique. These include high resolution tunnel spectroscopy with S-S junctions; the observation of the Josephson effect in tunnel regime with atomic size tips, which is the basis for scanning Josephson spectroscopy; and an experimental method to test surfaces with a tip whose density of states in the apical region can be changed, in a well controlled way, by means of external magnetic fields.

We acknowledge discussions with F. Guinea and A. Levanyuk and support from the MCyT, Spain (grant MAT-2001-1281-C02-0), and from the Comunidad Autónoma de Madrid, Spain (projects 07N/0039/2002 and 07N/0053/2002 ). The Laboratorio de Bajas Temperaturas is associated to the ICMM of the CSIC. The ESF Programme VORTEX is also acknowledged.